\documentclass[aps,prl,twocolumn,showpacs]{revtex4-1}
\usepackage{graphicx}
\begin{document}

\title{Microwave Surface Impedance Measurements of the Electronic State and Dissipation of Magnetic Vortices in Superconducting Iron-Based LiFeAs Single Crystals}

\author{
T. Okada$^{1,3}$, H. Takahashi$^{1,3}$, Y. Imai$^{1,3}$, K. Kitagawa$^{2,3}$, K. Matsubayashi$^{2,3}$, Y. Uwatoko$^{2,3}$ and A. Maeda$^{1,3}$
}
\affiliation{
$^{1}$Department of Basic Science, the University of Tokyo, Meguro, Tokyo 153-8902, Japan\\
$^{2}$Institute for Solid State Physics, the University of Tokyo, Kashiwa, Chiba 277-8581, Japan\\
$^{3}$Transformative Research Project on Iron Pnictides (TRIP), JST, Chiyoda, Tokyo 102-0075, Japan
}

\date{\today}

\begin{abstract}
LiFeAs is one of the iron-based superconductors having multiple gaps with the possible sign reversal.
To clarify how those novel natures affect the energy dissipation of magnetic vortices, we investigated the microwave surface impedance of LiFeAs single crystals under finite magnetic fields.
The flux-flow resistivity enhanced rapidly at low magnetic fields, which is similar to the case of MgB$_{2}$.
This is probably the consequence of the multiple-gap nature and the gap anisotoropy.
This suggest that the sign-reversal is not important for the flux-flow even for multiple-gap superconductors.
As for the electronic state, the vortex core of LiFeAs turned out to be ``moderately clean".
Furthermore, the mean free path inside the vortex core was much shorter than that outside, and was close to the core radius.
These results strongly suggest a process specific to the core boundary is important for a scattering mechanism inside the vortex core.
\end{abstract}


\maketitle
Since the discovery of LaFeAsO$_{1-x}$F$_{x}$ with $T_{c}=26$ K \cite{LaFeAsOF}, iron-based superconductors (SCs) have attracted lots of attention.
Because multiple bands contribute to the Fermi surfaces and the magnetic phase exists in the vicinity of the superconductive phase in the phase diagram, it is expected that the mechanism of superconductivity of iron-based SCs is different from that of conventional SCs.
New possibilities of superconducting gap structures based on the interband scattering, such as $s^{\pm}$-wave \cite{s+-Mazin,s+-Kuroki} and $s^{++}$-wave \cite{s++Onari,s++Yanagi}, were suggested theoretically.
Experimentally, although this issue is under a debate \cite{Review}, phase-sensitive experiments \cite{STM_11,JJ,STM_imp} suggested that $s^{\pm}$-state was realized in some materials of iron-based SCs.
It is of great interest what the electronic structure and dynamic properties of vortices in such novel class of SCs are.

As for conventional SCs, the quasiparticle (QP) excitation inside the vortex core has quantized energy levels with the spacing, $\mathit{\Delta}E\sim\Delta^{2}/E_{\mathrm{F}}\equiv\hbar\omega_{0}$, where $\Delta$ and $E_{\mathrm{F}}$ are the size of the superconducting gap and the Fermi energy, respectively, and with those width, $\delta E\sim\hbar/\tau_{\mathrm{core}}$, where $\tau_{\mathrm{core}}$ is the relaxation time of QPs inside the vortex core \cite{QPstate,ZEP}.
The ratio of these two energy scales, $\mathit{\Delta}E/\delta E\sim\omega_{0}\tau_{\mathrm{core}}$, is a barometer of the quantum nature of the electronic state inside the vortex core.
Depending on this number, we have three regimes as i) the dirty regime ($\omega_{0}\tau_{\mathrm{core}}\ll1$), i\hspace{-0.1em}i) the moderately clean regime ($\omega_{0}\tau_{\mathrm{core}}\sim1$) and i\hspace{-0.1em}i\hspace{-0.1em}i) the superclean regime ($\omega_{0}\tau_{\mathrm{core}}\gg1$).
It should be noted that $\omega_{0}\tau_{\mathrm{core}}$ is connected to the viscous drag coefficient, $\eta$, and the carrier density, $n$, as $\omega_{0}\tau_{\mathrm{core}}=\eta/n\pi\hbar$ \cite{vortex_core}.

According to Kopnin and Volovik (KV) \cite{KV}, the flux-flow resistivity of a single-gap SC, $\rho_{f}$, behaves in magnetic fields, $B$, as
\begin{equation}\label{eq:1}
\frac{\rho_{f}}{\rho_{\mathrm{n}}}\approx\frac{\Delta_{0}^2}{\langle{\Delta^{2}(\theta)\rangle}_{\mathrm{FS}}}\frac{B}{B_{c2}},\ \ \ (B\ll B_{c2})
\end{equation}
where $\rho_{\mathrm{n}},\ B_{c2},\ \Delta_{0}$ and $\langle\Delta^{2}(\theta)\rangle_{\mathrm{FS}}$ are the resistivity in the normal state, the upper critical field, the maximum size of the superconducting gap and the angular average of the superconducting gap on the Fermi surface, respectively.
This suggests that i) $\rho_{f}$ in low $B$ region increases linearly with $B$ and i\hspace{-0.1em}i) the gradient, $\alpha\equiv\Delta_{0}^2/\langle\Delta^{2}(\theta)\rangle_{\mathrm{FS}}$, becomes larger than unity when $\Delta(\theta)$ depends on the angle $\theta$.
In fact, for an isotropic gap case, the Bardeen-Stephen (BS) theory \cite{BS} obviously obeys Eq.(\ref{eq:1}).
On the other hand, in nodal and modulated gap case, an enhancement with $\alpha>1$ at low $B$ region has been also observed experimentally \cite{UPt,YBCO,TBCO,YNBC}.
This also suggests that the so-called ``Volovik effect" (the effect of the Doppler shift on QPs disperse caused by the circulating supercurrents) is not important for the flux-flow in low $B$ region, although it succeeded to explain $B$ dependences of the specific heat and the thermal conductivity.
As for the 2-band $s^{++}$-wave SCs, such as MgB$_{2}$ and Y$_{2}$C$_{3}$, a rapid enhancement of $\rho_{f}(B)$ was observed \cite{MgB2,Y2C3}.
This can be interpreted as the superposition of two linear $B$ dependences corresponding to two bands \cite{Goryo}.
Thus, $\rho_{f}(B)$ reflects the superconducting gap structure and its symmetry.
Therefore, it is very interesting how the flux-flow resistivity of the novel class of SCs behaves as a function of $B$.
However, the flux-flow of such novel SCs has not been investigated at all both theoretically and experimentally.
Thus, it is a great challenge to investigate the flux-flow of iron-based SCs.

We focus on a 111 material, LiFeAs.
This material has the highest $T_{c}$ of 18 K \cite{Tc18K} among stoichiometric iron-based SCs, and single crystals with high quality (residual resistivity ratio ($RRR$)$\sim$50) can be obtained.
The band calculation \cite{Band111} suggested that Fermi surfaces consist of two hole-like and two electron-like pockets around $\Gamma$-points and M-points, respectively.
Nodeless multiple superconducting gaps were observed by an angle-resolved photoemission spectroscopy (ARPES) \cite{ARPES111,ARPES1112} and a specific heat measurement \cite{C}, superfluid-density data \cite{lambda111_Kim,lambda111_Imai} showed that LiFeAs has nodeless multiple-gap structure.
In addition to the phase sensitive experiment in Li-111 \cite{STM_imp}, the electrical conductivity, $\sigma_{1}$ \cite{lambda111_Imai}, estimated from the microwave surface impedance and the nuclear spin-lattice relaxation rate, $1/T_{1}$ \cite{NMR111}, do not show the so-called ``coherence peak" below $T_{c}$.
These strongly suggest that LiFeAs has the $s^{\pm}$-wave gap structure.
Therefore, we can stand for the standpoint that Li-111 is an $s^{\pm}$-SC.

In this paper, we report the surface impedance of LiFeAs single crystals under finite magnetic fields, and discuss the electronic state inside the vortex core.
It was clarified that the field dependence of the flux-flow of $s^{\pm}$-state is similar to that of $s^{++}$-state, and that the vortex core of LiFeAs is moderately clean.
The estimated mean free path of QPs inside the vortex core was found to be much shorter than that outside, and comparable to the core radius.
This suggest that the mechanism characteristic to the core boundary plays an important role in the dissipative process inside the vortex core. 

LiFeAs single crystals were grown by a self-flux method \cite{lambda111_Imai} and were cleaved under Ar atmosphere in a glove box.
Typical size of sample was $0.5\times0.5\times0.2\ \mathrm{mm}^{3}$, and the demagnetization coefficient estimated under ellipsoidal approximation was about 0.58.
These were of very high quality with $RRR\equiv\rho_{\mathrm{dc}}(300\ \mathrm{K})/\rho_{\mathrm{dc}}(T_{c})\sim45$, and the dc resistivity behaved as $\rho_{\mathrm{dc}}(T>T_{c})=\rho_{0}+AT^{2}\ (\rho_{0}\approx30\ \mu\Omega\mathrm{cm},\ A\approx6.5\times10^{-2}\ \mu\Omega\mathrm{cm}/\mathrm{K}^{2})$, which is typical of the Fermi liquid dominated by the electron-electron scattering.
Since LiFeAs is moisture/atmosphere sensitive, samples were covered with Apiezon N grease during the measurement.
We confirmed that Apiezon N grease does not affect results discussed below in a different comparative experiment.

The microwave surface impedance was measured by using a cavity perturbation technique \cite{CPM} with a cylindrical oxygen-free Cu cavity resonator operated at $\omega/2\pi\sim19$ GHz in the TE$_{011}$ mode.
The $Q$-factor was $Q\gtrsim6\times10^{4}$, and the filling factor of samples was about $6\times10^{-6}$.
Both the external magnetic field up to 8 T and the microwave magnetic field were applied parallel to the $c$-axis.
Therefore, we investigated the in-plane vortex motion.

The surface impedance, $Z_{\mathrm{s}}=R_{\mathrm{s}}-iX_{\mathrm{s}}$ ($R_{\mathrm{s}}$ and $X_{\mathrm{s}}$ are the surface resistance and the surface reactance, respectively), is related to the resonant frequencies, $\omega_{s}/2\pi$ and $\omega_{b}/2\pi$, and the $Q$-factors, $Q_{s}$ and $Q_{b}$, as $R_{\mathrm{s}}=G\left(1/2Q_{s}-1/2Q_{b}\right),\ X_{\mathrm{s}}=G\left(1-\omega_{s}/\omega_{b}\right)+C$, where subscripts $s$ and $b$ represent the values measured with- and without the sample, respectively, and $G, C$ are constants determined by the size and the shapes of the sample and the resonator.
The magnitudes of $R_{\mathrm{s}}$ and $X_{\mathrm{s}}$ are obtained by assuming the Hagen-Rubens relation, $R_{\mathrm{s}}=X_{\mathrm{s}}=\sqrt{\mu_{0}\omega\rho_{\mathrm{dc}}/2}$, in the normal state.

$Z_{\mathrm{s}}$ in the mixed state was calculated by Coffey and Clem (CC) \cite{CC}.
Their calculation is based on the equation of motion of the massless vortex, $\eta\dot{\mbox{\boldmath $u$}}+\kappa\mbox{\boldmath $u$}=\Phi_{0}\mbox{\boldmath $J$}\times\hat{\mbox{\boldmath $z$}}+\mbox{\boldmath $f$}(t)$, where $\mbox{\boldmath $u$}$ is the displacement of a vortex, $\kappa$ is the pinning force constant, $\Phi_{0}=h/2e=2.07\times10^{-15}$ Wb is the flux quantum, \mbox{\boldmath $J$} is the transport current density and $\hat{\mbox{\boldmath $z$}}$ is the unit vector in the applied field direction.
The effect of thermal fluctuations and the Hall effect are effectively included in random force, $\mbox{\boldmath $f$}(t)$, and $\eta$, respectively for circulating microwave currents.
At low temperature, the flux-creep contribution becomes negligibly small and the CC model leads to the relation
\begin{equation}\label{eq:2}
	Z_{\mathrm{s}}=-i\mu_{0}\omega\sqrt{\frac{\lambda_{\mathrm{L}}^{2}+\frac{1}{\mu_{0}\omega}\rho_{f}(1-i\frac{\omega_{\mathrm{cr}}}{\omega})^{-1}}{1+is}},
\end{equation}
where $\lambda_{\mathrm{L}}$ is the London penetration depth, and $\omega_{\mathrm{cr}}/2\pi$ is the crossover frequency characterizing the crossover between reactive- and resistive response, and $s=\mu_{0}\omega\lambda_{\mathrm{L}}^{2}/\rho_{\mathrm{n}}$ which represents the normal-fluid contribution.
One can assume that $s$ to be negligible at low temperatures.
Consequently, we obtain $\omega_{\mathrm{cr}}$ and $\rho_{f}$ from experimental data of $R_{\mathrm{s}}$ and $X_{\mathrm{s}}$, by solving Eq.(\ref{eq:2}).
 
\begin{figure}[h]
	\includegraphics[width=6.5cm]{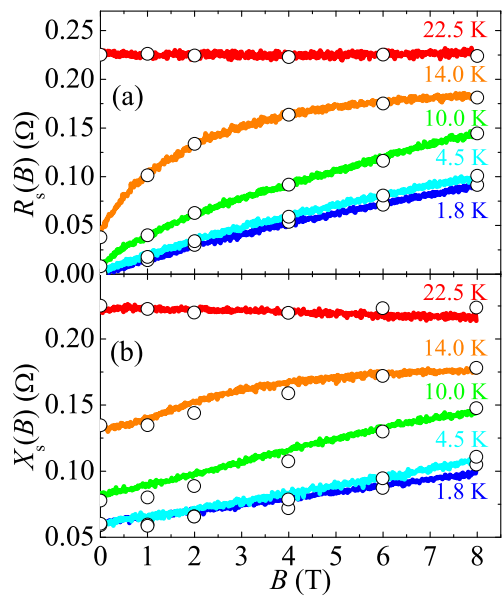}
	\caption{\label{f1}(Color online) The magnetic field dependence of (a) the surface resistance, $R_{\mathrm{s}}$, and (b) the surface reactance, $X_{\mathrm{s}}$, of a LiFeAs single crystal at 19 GHz up to 8 T at various temperatures.
The curves and the open circles represent the data taken in the swept magnetic field (fixed temperatures) and in the swept temperature (fixed magnetic fields), respectively.}
\end{figure}

Figure \ref{f1} shows the magnetic field dependence of $Z_{\mathrm{s}}$ at various temperatures.
Good agreement between temperature swept data and magnetic-field swept data represents that the magnetic field penetrates uniformly in the sample.
With increasing magnetic field, both $R_{\mathrm{s}}$ and $X_{\mathrm{s}}$ increase monotonically.
In particular, $R_{\mathrm{s}}$ shows a convex upward behavior.
We determine the zero-field superconducting transition temperature, $T_{c}^{\mathrm{onset}}=17$ K, from the temperature dependence of $X_{\mathrm{s}}$ in zero magnetic field, which is in good agreement with the previously reported number in the same batch \cite{lambda111_Imai}.

The crossover frequency of $\omega_{\mathrm{cr}}/2\pi\approx3$ GHz obtained is larger than that of conventional SCs ($\approx100$ MHz) \cite{micro} but smaller than that of copper-oxide SCs by one order of magnitude \cite{f0_Golosovsky,f0_Revenaz}.
A similar value of $\omega_{\mathrm{cr}}$ has been reported in a 1111-type polycrystal ($\approx6$ GHz) \cite{Zs1111}.
The tendency that $\omega_{\mathrm{cr}}$ becomes small at high temperatures is consistent with a general description that the thermal fluctuation decreases the pinning force.
 
\begin{figure}[h]
	\includegraphics[width=6cm]{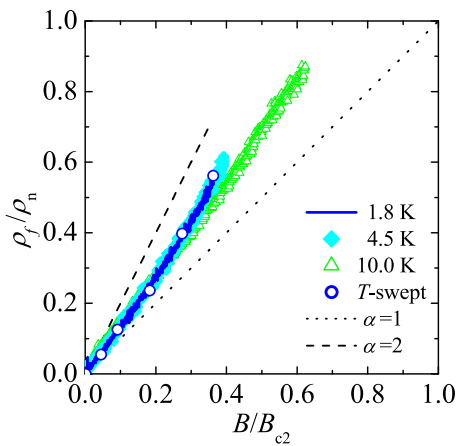}
	\caption{\label{f2}(Color online) The magnetic field dependence of the flux-flow resistivity $\rho_{f}(B)$ of the LiFeAs single crystal at several temperatures.
The blue open circle is $\rho_{f}(B)$ at $T$=1.8 K obtained from temperature swept data.
The gradient, $\alpha$, expected in $d$-wave (with lines of node) SCs ($\alpha\approx2$) and in conventional $s$-wave SCs ($\alpha=1$) are also shown as dashed- and dotted lines, respectively.}
\end{figure}

Figure \ref{f2} shows the normalized flux-flow resistivity as a function of the normalized magnetic field.
The flux-flow resistivity of LiFeAs single crystals increased linearly with $B$, suggesting that the KV model is appropriate even for this material.
As for the gradient, $\alpha$ of LiFeAs is larger than that of the conventional $s$-wave case ($\alpha=1$) and smaller than that of the $d$-wave (with lines of node) case ($\alpha\approx2$).
This enhancement of $\rho_{f}(B)$ may be derived from one or both of two origins.
First possible origin is based on the multiple-band nature.
As for the 2-bands SCs, such as MgB$_{2}$ and Y$_{2}$C$_{3}$, the superposition of two linear dependences corresponding to two bands cause the flux-flow resistivity enhanced rapidly at low $B$ \cite{MgB2,Y2C3}.
We can speculate that the 5-bands nature of LiFeAs probably induces the similar tendency.
Second possibility is based on the gap anisotropy.
Recent ARPES data suggests that some of the superconducting gaps have obvious 4-fold angle dependences \cite{ARPES1112}.
Based on the KV model, this angle dependence of the superconducting gap will make the gradient of $\rho_{f}(B)$ larger than unity ($\alpha>1$).
In any case, the magnetic field dependence of $\rho_{f}$ of LiFeAs are very similar to that of MgB$_{2}$, implying that the $s^{\pm}$-wave SC behaves essentially similarly to the $s^{++}$-wave SC so far as the flux-flow is concerned.
The insensitivity of the flux-flow to the sign change for single-gap SCs has been already known for single-gap SCs; although the anisotropic $s$-wave SC and the $d$-wave SC differ from each other in the sign change of the order parameter, $\rho_{f}$ of both SCs show the $B$-linear dependence with $\alpha>1$ \cite{YBCO,TBCO,YNBC}.
Our present result shows that the insensitivity shown up in the flux-flow is applicable also for multiple-gap SCs.

\begin{figure}[h]
	\includegraphics[width=7cm]{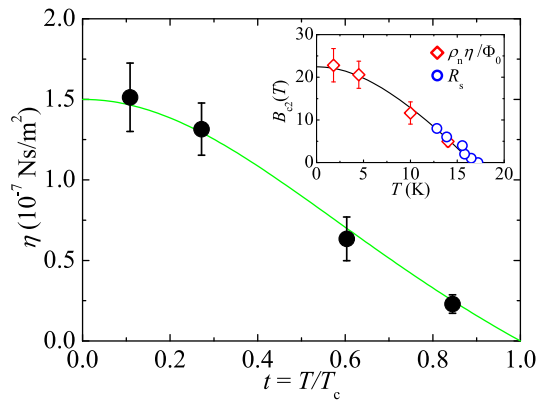}
	\caption{\label{f3}(Color online) The temperature dependence of the viscous drag coefficient, $\eta=\Phi_{0}B/\rho_{f}$.
The green solid line is the expectation in the GL theory, $\eta(t)=\eta(0)(1-t^{2})/(1+t^{2})$.
The inset shows the temperature dependence of $B_{c2}$ obtained from $T_{c}^{\mathrm{onset}}$ from the temperature dependence of $R_{\mathrm{s}}$ (blue open circle), and that calculated from $\rho_{\mathrm{n}}$ and $\eta$ (red open diamond). The solid line is eye guide.}
\end{figure}

Figure \ref{f3} shows the temperature dependence of the viscous drag coefficient, $\eta=\Phi_{0}B/\rho_{f}$.
$\eta$ is well fitted by the expected temperature dependence in the Ginzburg-Landau (GL) theory, $\eta(T)=\eta(0)[1-(T/T_{c})^{2}]/[1+(T/T_{c})^{2}]$.
From the fitting, we obtain $\eta(0)=(1.5\pm0.2)\times10^{-7}$ Ns/m$^{2}$.
We can estimate the upper critical field as $B_{c2}(T)=\rho_{\mathrm{n}}(T)\eta(T)/\Phi_{0}$,
where $\rho_{\mathrm{n}}(T)=\rho_{0}+AT^{2}$ is extrapolated $\rho_{\mathrm{dc}}(T>T_{c})$ to the temperature regions $T<T_{c}$.
The result is shown in the inset.
We obtain $B_{c2}(0)=22\pm4$ T.
Similar numbers were reported previously in the same material \cite{Hc2_Kurita,Hc2_Zhang,Hc2_Cho,Hc2_Tanatar,Hc2_Heyer,Hc2_Song}.
Considering the moisture/atmosphere sensitive nature of LiFeAs and the difference of $RRR$ values among these crystals, we consider that it is within the range of individual differences.

\begin{figure}[h]
	\includegraphics[width=8cm]{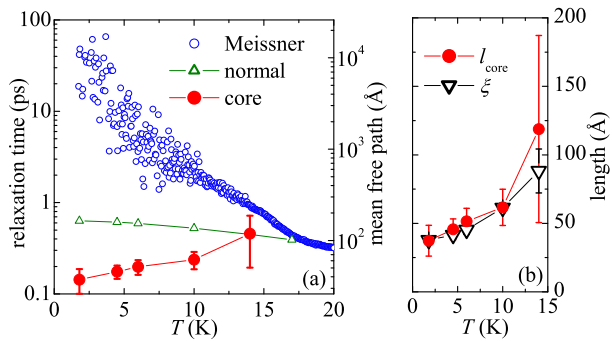}
	\caption{\label{f4}(Color online) (a) Temperature dependences of relaxation times and mean free paths of QPs in several states.
Symbols are those outside the vortex core (blue open circle), in the normal state (green open triangle) and inside the vortex core (red solid circle), respectively.
(b) Temperature dependences of the mfp inside the vortex core (red circle) and the coherence length calculated from $B_{c2}(T)$ (black triangle).}
\end{figure}

In figure \ref{f4}, we discuss the relaxation time and the mean free path (mfp) of QPs inside the vortex core.
Since $\lambda_{\mathrm{L}}^{-2}(T)=\mu_{0}e^{2}n_{s}(T)/m^{\ast}$, using the value $m^{\ast}/m_{0}=5.2-6.3$ ($m_{0}$ is the free-electron mass) \cite{dHvA,AsP} and $\lambda_{\mathrm{L}}(0)=X_{\mathrm{s}}(0)/\mu_{0}\omega\sim390$ nm, we estimate the carrier density $n\approx n_{s}(0)=(9.6-11.7)\times10^{20}\ \mathrm{cm}^{-3}$, which gives $\omega_{0}\tau_{\mathrm{core}}=0.4-0.5$.
This shows that the vortex core of LiFeAs is in the moderately clean regime.
Furthermore, by using the number $-\hbar\omega_{0}/2=-0.9$ meV observed in a recent scanning tunneling microscopy/spectroscopy (STM/STS) study \cite{STM_111}, we obtain the relaxation time of QPs inside the vortex core, $\tau_{\mathrm{core}}(1.8\ \mathrm{K})\approx0.15$ ps.
This value is quite different from that outside ($\approx10$ ps) \cite{lambda111_Imai}, and is even smaller than that in the normal state ($\approx0.6$ ps).
These are shown in Figure \ref{f4}(a).
From the relaxation time, we found that the mfp of QPs inside the vortex core to be $l_{\mathrm{core}}=v_{\mathrm{F}}\tau_{\mathrm{core}}\approx40$ \AA, where $v_{\mathrm{F}}\approx2.6\times10^{4}$ m/s is the Fermi velocity, which is estimated from STM/STS \cite{STM_111} and ARPES data \cite{ARPES111,ARPES1112}.
Again, this value is much shorter than that outside the core, $l_{\mathrm{Meissner}}$.
In particular, as shown in Figure \ref{f4}(b), $l_{\mathrm{core}}$ is comparable to the coherence length, $\xi$, estimated from $B_{c2}$.
We checked the repeatability in another single crystal of LiFeAs, and the results were consistent with those described above.
In addition, we performed the same measurements in LiFe(As,P) single crystals, which was at most 3 \% P-substituted, and we obtained the similar results.

The short mfp of QPs inside the vortex core was also observed in many copper-oxide SCs, such as YBa$_{2}$Cu$_{3}$O$_{7-x}$, Bi$_{2}$Sr$_{2}$CaCu$_{2}$O$_{y}$ and La$_{2-x}$Sr$_{x}$CuO$_{4}$ \cite{YBCO_lcore,BSCCO,LSCO}.
In these cuprate, the mfp inside the vortex core is also much shorter than that outside and rather close to the core radius, $l_{\mathrm{M}}\gg l_{\mathrm{core}}\sim\xi$.
Similarly, in Y$_{2}$C$_{3}$ \cite{Y2C3}, which is one of the 2-gap SCs with isotropic $s$-wave, the mfp inside the vortex core is limited to the coherence length, $l_{\mathrm{core}}\lesssim\xi$.
It is surprising that similar tendency was observed among many different SCs with different gap structures, pairing mechanisms and electronic structures.
Since the relation, $l_{\mathrm{core}}\sim\xi$, was obtained, one can consider that a scattering process which is specific to the core boundary contributes to the additional dissipation in the vortex core as was originally considered by Nozi$\grave{\textrm{e}}$res and Vinen for clean SCs \cite{NV}.
Indeed, Eschrig $et\ al.$ \cite{CollectiveMode} discussed that the Andreev reflection at the core boundary is crucial even in the flux-flow of moderately clean SCs, and theoretically showed that there is extra energy dissipation at low frequencies because of the presence of a collective mode.
However, it is not yet clear whether this mechanism can explain the large dissipation observed in our experiments quantitatively at present.
Systematic study of the frequency dependence of the in-core dissipation will clarify the validity of Eschrig's model.
On the other hand, according to Tinkham \cite{Tinkham} and Nozi$\grave{\mathrm{e}}$res-Vinen-Warren \cite{NV,VW}, the relaxation time $\tau_{\mathrm{gap}}=\hbar/\Delta_{0}$ which is characteristic of the moving vortex, has been considered.
For LiFeAs, $\tau_{\mathrm{gap}}=0.2$ ps is comparable to obtained $\tau_{\mathrm{core}}$.
In order to clarify the validity of these models, studies of the gap-size dependence of $\tau_{\mathrm{core}}$ is needed.

In conclusion, we investigated the microwave surface impedance of LiFeAs single crystals under finite magnetic fields.
The magnetic field dependence of the flux-flow resistivity of new class of superconductors having multiple gaps with the possible sign reversal became clear.
The flux-flow resistivity increased linearly with the magnetic field, as was suggested by Kopnin-Volovik.
Particularly, the gradient at low fields was larger (smaller) than that of conventional $s$-wave superconductors ($d$-wave superconductors with lines of node).
This is probably the consequence of the multiple-gap nature and/or the gap anisotropy.
This also suggests that the flux-flow resistivity is insensitive to the sign reversal of the order parameter on different Fermi surfaces.
As for the electronic state, the vortex core of LiFeAs was estimated to be the moderately clean.
The mean free path of quasiparticles inside the vortex core was much shorter than that outside, and comparable to the core radius, suggesting the importance of the Andreev reflection at the core boundary.
Such a tendency was observed also in many other superconductors, and systematic studies will clarify the dissipative mechanism inside the vortex core.

\begin{acknowledgments}
We thank Tetsuo Hanaguri for showing us many unpublished data and also for fruitful discussions.
We also thank Masashi Takigawa for providing us LiFeAs single crystals, and Yusuke Kato for valuable comments.\end{acknowledgments}

\providecommand{\noopsort}[1]{}\providecommand{\singleletter}[1]{#1}
\end{document}